%% file: proximity-revised-prb.tex
\documentclass[twocolumn,floatfix,superscriptaddress]{revtex4-1}

\usepackage{amsmath,amssymb,wscmath}
\usepackage{color}
\usepackage[pdftex]{graphicx, hyperref}
\hypersetup{colorlinks = true, urlcolor = blue, linkcolor = blue, citecolor = blue}

\begin{document}
\title{Effects of large induced superconducting gap on semiconductor Majorana nanowires}
\author{William S. Cole}
\author{S. Das Sarma}
\affiliation{Condensed Matter Theory Center and Joint Quantum Institute,
Department of Physics, University of Maryland, College Park, MD 20742, USA}
\author{Tudor D. Stanescu}
\affiliation{Department of Physics and Astronomy, West Virginia University, Morgantown, WV 26506, USA}
\begin{abstract}
With the recent achievement of extremely high-quality epitaxial interfaces between InAs nanowires and superconducting Al shells with strong superconductor-semiconductor tunnel coupling, a new regime of proximity-induced superconductivity in semiconductors can be explored where the induced gap may be similar in value to the bulk Al gap (\emph{large} gap) with negligible subgap conductance (\emph{hard} gap). We propose several experimentally relevant consequences of this large-gap strong-coupling regime for tunneling experiments, and we comment on the prospects of this regime for topological superconductivity. In particular, we show that the advantages of having a strong spin-orbit coupling and a large spin $g$-factor in the semiconductor nanowire may both be compromised in this strongly coupled limit, and somewhat weaker interface tunneling may be necessary for achieving optimal proximity superconductivity in the semiconductor nanowire. We derive a minimal, generic theory for the strong-coupling hard-gap regime obtaining good qualitative agreement with the experiment and pointing out future directions for further progress toward Majorana nanowires in hybrid semiconductor-superconductor structures.
\end{abstract}
\date{\today}
\maketitle

\section{Introduction}
Majorana zero-energy modes (MZMs) -- whose creation and annihilation operators are identical, as envisioned by Ettore Majorana in the context of obtaining a real version of the Dirac equation 80 years ago \cite{majorana-original} -- are expected to exist as localized subgap excitations of spinless p-wave superconductors \cite{readgreen-MZM, kitaev-MZM}. In two dimensions, such MZMs would manifest non-Abelian braiding statistics which can be used for fault-tolerant topological quantum computation \cite{nayak-rmp} . Although these theoretical developments attracted a great deal of attention, no naturally occurring, low-dimensional, spinless p-wave superconductor has yet been discovered (though Sr$_2$RuO$_4$ seems to come close \cite{sds-sr2ruo4, budakian-sr2ruo4}).  A series of theoretical papers \cite{fukane-prl, zhang-pwave-prl, sato-pwave-prl, SauSM_PRL, SauSM_PRB, LutchynSM_PRL, OregSM_PRL, AliceaSM_PRB} suggested that if spinless p-wave superconductors (``topological superconductors" for our purposes in this article, being defined here as a system that can host non-Abelian MZMs) do not exist in nature, perhaps they can be artificially engineered from simpler ingredients. One specific idea that attracted wide-spread experimental attention is the hybrid semiconductor-superconductor (SM-SC) nanowire architecture. In this architecture, a spin-orbit-coupled semiconductor nanowire (e.g. InAs, InSb) with large Zeeman spin-splitting is proximity-coupled to an ordinary s-wave superconductor (e.g. Nb, Al). It was shown in \cite{LutchynSM_PRL, OregSM_PRL} that such a semiconductor nanowire becomes a topological superconductor if the proximity-induced superconducting gap opens up precisely in the ``helical gap" created in the nanowire band structure from the combined effect of strong spin-orbit coupling and large Zeeman splitting. Immediately following these theoretical predictions, an experiment \cite{mourik-expt} was reported showing signatures of MZMs in an applied magnetic field for InSb nanowires coupled to superconducting NbTiN.  This experimental finding was quickly verified in several following measurements \cite{mourik-expt, das-expt, deng-expt, finck-expt, churchill-expt} using InSb or InAs nanowires and Nb or Al as the parent superconductor, creating tremendous excitement.  The excitement and the activity in this subject have grown during the last few years, and several review articles have appeared in the literature \cite{MZMandTQC, beenakker-review, alicea-review, flensberg-review, stanescu-review, franz-review}.

\begin{figure}[t]
\begin{center}
\includegraphics[width=0.4\textwidth]{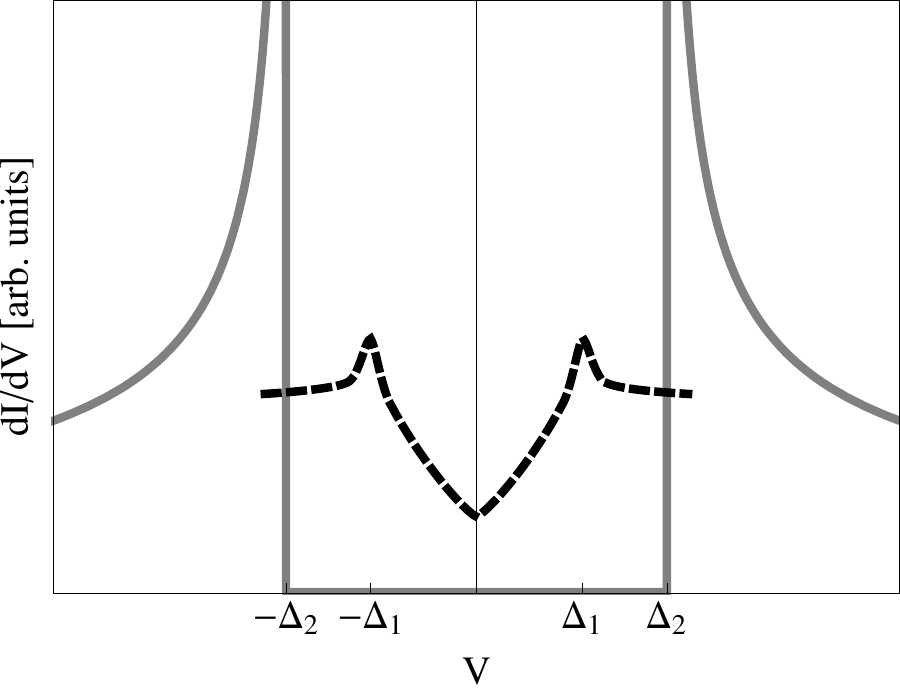}
\caption[hard vs. soft gap schematic]{Schematic illustration of the differences between hard and soft gaps in the context of proximity superconductivity. The soft gap tunnel conductance has several features common across several experiments: a clear but broad maximum (at $\pm \Delta_1$) associated with the coherence peak, and a v-shaped subgap conductance with a minimum at zero energy where the conductance is still a substantial fraction of the normal state conductance. For comparison, we show a characteristic hard gap tunnel conductance, similar to that seen in \cite{cm-hardgap}. Here there are very sharp coherence peaks at the induced gap scale $\pm \Delta_2$, with negligible subgap conductance. The measured zero-field zero-bias conductance scales in the hard gap \cite{cm-hardgap} and the soft gap \cite{mourik-expt} systems differ by an order of magnitude (being 0.03 and 0.3 roughly in the units of $e^2/h$ for Refs.~\cite{cm-hardgap} and \cite{mourik-expt} respectively)}
\label{fig:hardsoftgap}
\end{center}
\end{figure}

Despite all this activity, there remain issues and questions about the experimental findings. The most notable problem, which is the subject matter of the present work, has been the absence of an observable ``hard gap" in the proximitized semiconductor's density of states, as determined by differential conductance measurements \cite{mourik-expt, das-expt, deng-expt, finck-expt, churchill-expt} (see Fig.~\ref{fig:hardsoftgap}).  In particular, the induced superconducting gap in the nanowire appears to be universally \emph{soft} in all of these experiments with the subgap conductance being at best a factor of 4 smaller than the above-gap normal conductance, casting serious doubt on the whole question of whether there is even any superconductivity in the nanowire at all. Various possible reasons for the soft proximity-induced gap in the nanowire have been put forward \cite{takei, tudor-softgap, setiawan, sun-btk}, but there have also been suggestions \cite{altland-disorder, lee-disorder, beenakker-disorder, sau-disorder, kells-confinement, lee-kondo} that perhaps the observed experimental signatures have nothing to do with MZMs per se but arise from a combination of spin-orbit coupling, particle-hole symmetry, and Zeeman splitting and do not require a fully gapped wire at all. In addition to these serious reservations regarding even the possible existence of MZMs, a soft gap would be a serious problem for the non-Abelian braiding properties of MZMs, if present, as the Majorana would simply decay into these ordinary fermionic states.

The soft gap problem is therefore a central issue to the whole experimental progress of the subject of Majorana nanowires and not just a technical nuisance. This is the context for an important new experiment from the Copenhagen group \cite{cm-hardgap} which has reported the observation of a \emph{large} and \emph{hard} superconducting proximity gap in InAs nanowires, induced by Al as the parent superconductor using clever epitaxial core-shell nanowire growth techniques \cite{krogstrup-natmat}.  The proximity gap in the InAs nanowires in this Copenhagen experiment looks essentially identical to the bulk superconducting gap of the parent Al and therefore all the reservations discussed above with respect to the soft gap in Majorana nanowires seem to disappear in one stroke by this potential breakthrough experiment.

In this work we provide careful theoretical analyses and simple numerical simulations motivated by this Copenhagen experiment. Using a generic, minimal model for proximity-induced superconductivity in the hybrid system, we uncover some new general principles; for example, while a soft gap is always detrimental, we point out that this experiment may have actually solved the problem \emph{too} well by entering a regime of strong semiconductor-superconductor coupling, necessary for the large gap they observed. A large gap is desirable for a variety of reasons, but we demonstrate that such strong coupling produces its own set of rather serious issues which might compromise the study of MZMs in these new hard-gap nanowires.

Although we focus on the SM-SC hybrid structure of Refs.~\cite{cm-hardgap,krogstrup-natmat} (since this is the platform reporting the observation of a hard induced SC gap) our minimal theory is applicable in principle to all hybrid superconducting structures (e.g. both one- and two-dimensional SM-SC structures, ferromagnetic chains on superconductors, topological insulators on superconductors, quantum spin hall layers on superconductors, etc.) manifesting proximity effect -- the effects that we consider should be accounted for in interpreting any hybrid SC system.


\section{Questions for theory}
While the experiment by Chang et al. \cite{cm-hardgap} appears to be a breakthrough in the study of SM-SC Majorana nanowires, a conundrum arises because the tunneling data in Ref.~\cite{cm-hardgap} appears to be essentially identical to that expected for the parent SC Al, raising the possibility that the unique properties of the SM system are completely overwhelmed by the parent SC. In fact, some of the data of Ref. \cite{cm-hardgap} are not inconsistent with an experiment actually probing only the superconducting properties of the Al shell itself (rather than the InAs core) without the manifestation of any proximity effect. We will further elucidate this point in Sec.~\ref{sec:maki}.

This sets the broad motivation for our work: how relevant is the strong-coupling regime (i.e., with a large induced gap, relative to the parent SC gap) for the realization and observation of MZM physics? The properties of the induced superconducting state should, of course, depend critically on the strength of the effective SM-SC coupling; in turn, this coupling is determined by the specific materials and by the quality and area of the interface and many other microscopic details beyond the scope of the present work. Nonetheless, a thorough understanding of the dependence of the proximity-induced SC state on the SM-SC coupling strength presents a key challenge for theory. It is critical at first to obtain a clear \emph{qualitative} understanding of the specific experimental signatures of proximity-induced features at strong coupling, e.g., in charge tunneling experiments, which have to date been the most widely used probe (and is indeed the technique used by Chang et al. \cite{cm-hardgap} to study the InAs/Al SM-SC system through the usual normal-superconductor (NS) tunneling spectroscopy).

What do we learn by measuring the tunneling conductance in a strong-coupling SM-SC system? For an NS junction in the weak tunneling (i.e., strong barrier) regime, one would expect $dI/dV$ to be proportional to the local density of states (LDOS) on the SC side of the (sharp) junction. For a strong but \emph{smooth} barrier, such as that created by a gate potential, it is natural to assume that $dI/dV$ actually ``samples" the LDOS within some finite region of the SC. If the SC is homogeneous, there is no qualitative difference between the sharp and the smooth barriers. However, when tunneling into a proximity-coupled hybrid structure, the region sampled by $dI/dV$ may extend into \emph{both} components, so that the differential conductance will look, at least qualitatively, like the sum of LDOS contributions near the end of the SM wire and LDOS contribution from the adjacent region of the bulk superconductor. We note that previous calculations of LDOS in Majorana hybrid structures have only considered the SM wire contributions, as they either completely ignore the bulk SC (by considering an induced pair potential $\Delta_{\rm ind}$ as an input parameter), or replace the superconductor by an interface self-energy contribution. This is perfectly reasonable \cite{jay-proxeffect} when the SM states of interest are much lower in energy than the bare gap of the parent SC, but this is not the case for Ref.~\cite{cm-hardgap}. The tunneling conductance gap there, ostensibly measured in an InAs wire, appears to be nearly \emph{identical} to the gap of the SC Al shell. What is the nature of the proximity effect in such a scenario? What role, if any, are SM states playing in the conductance? Why does the observed proximity induced tunneling gap in the semiconductor nanowire appear to be essentially identical to that of the parent bulk metallic superconductor?  These are the questions we address theoretically in our work, which then lead naturally to suggestions on what one will have to do in fabricating the next generation core-shell epitaxial hard gap nanowires in order to optimize Majorana properties.

In the following sections we make several observations and predictions to guide future tunneling experiments. First, we elaborate more on the motivation for our work by analyzing the magnetic field dependence of the tunneling DOS reported in Ref.~\cite{cm-hardgap}, which is remarkably consistent with theoretical expectations for a pure Al wire; this suggests that the contribution of InAs states to the conductance of the device is subdominant (in fact, negligible in the reported full-shell data of Ref.~\cite{cm-hardgap}). This puzzle in hand, we then demonstrate with a simplified but explicit model that for strong SM-SC coupling the originally subgap InAs states are driven to the bare bulk SC gap, where they are then hidden under the LDOS contribution from the SC (rendering the semiconductor contribution invisible in the tunneling data). We point out other interesting generic features in the tunneling LDOS that should be observable if the SC contribution could be subtracted out (or otherwise suppressed), mentioning that some features of this qualitative prediction are already observable in the half-shell nanowire results presented in Ref.~\cite{cm-hardgap} where \emph{presumably} the SM-SC coupling is weaker than in the full-shell system leading to the appearance of weak features associated with the SM in the SC gap. Finally, we describe a minimal theoretical model which emphasizes an essential competition: having a hard and large SC proximity gap versus discerning the SM properties essential for creating the topological phase. In driving the subgap wire states up to the bare gap scale, other key parameters, for example the effective $g$ factor of the wire, get substantially renormalized. As a result, for sufficiently strong coupling, the conditions for topological superconductivity are drastically modified and may be challenging (perhaps even impossible) to attain. This all leads to the inevitable conclusion that extreme strong-coupling superconducting hybrid structures should be avoided in studying topological Majorana physics.


\section{Magnetic field dependence of the SC density of states}\label{sec:maki}
The physical consequences of Cooper pair breaking induced by an orbital magnetic field in a small, disordered superconductor were first worked out by Maki \cite{maki-ptp}. Since the formal details of the calculations can be found in several classic references \cite{maki-ptp, maki-book, skalski}, we provide here only a minimal recap of the ingredients and our calculated results of the theory relevant to the SM-SC system of Ref.~\cite{cm-hardgap}.

The generic starting point is the pair-breaking-induced renormalization of the bare frequency $\omega \rightarrow \tilde{\omega}$ and mean-field pair potential $\Delta \rightarrow \tilde{\Delta}$ in the superconductor Green's function. The expression relating the bare and renormalized values are particularly concise in terms of a parameter $u \equiv \tilde{\omega}/\tilde{\Delta}$,
\be\label{eq:maki_eq1}
\frac{\omega}{\Delta} = u \left( 1 - \frac{\alpha}{\Delta} \frac{1}{\sqrt{1-u^2}} \right)
\ee
where $\alpha \in \left[ 0, \Delta_{00}/2 \right]$ is a tunable parameter describing the strength of the pair-breaking effect, and $\Delta_{00}$ is the pair potential at zero temperature in the absence of pair breaking. $\Delta$ itself must still be calculated self-consistently as a function of $\alpha$, and the upper bound given on $\alpha$ is determined from the value where $\Delta(\alpha)$ vanishes, and superconductivity along with it. For the present consideration of Ref.~\cite{cm-hardgap} in an applied magnetic field $B$, $\alpha \propto B^2$ \cite{maki-book}. Then, given a critical field $B_c$ above which superconductivity vanishes, we can write $\frac{2\alpha}{\Delta_{00}} = \left(\frac{B}{B_c}\right)^2$.

We want a quantitative understanding of how we expect pair breaking to impact the tunneling DOS. In this formalism, the DOS is expressed as
\be\label{eq:maki_eq2}
\frac{N_S(\omega)}{N_N(0)} = \Im \frac{u}{\sqrt{1-u^2}}
\ee
where $N_N(0)$ is the density of states at the Fermi energy in the normal state.

\begin{figure}[t]
\begin{center}
\includegraphics[width=0.48\textwidth]{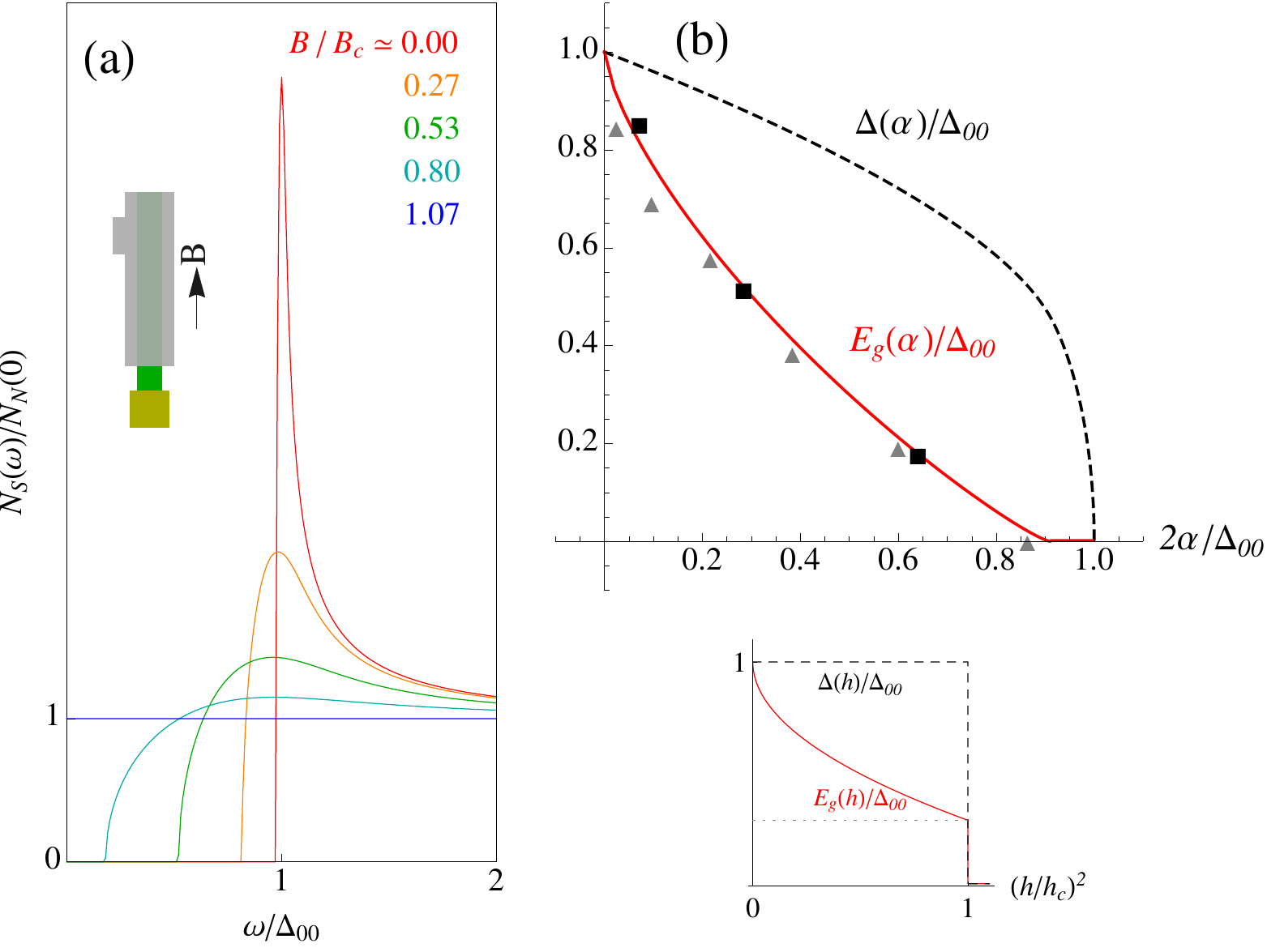}
\caption[Maki summary]{(Color online)  (a) Expected tunneling densities of states in the Maki formalism at values $\alpha$ corresponding to the values $B/B_c$ in Fig.~4a of \cite{cm-hardgap}. Inset is a schematic drawing of the tunneling experiment geometry from \cite{cm-hardgap}. An InAs core (green) is symmetrically coated with Al (gray); part of the coating is etched away and the wire is contacted by a normal metal (yellow). An additional superconducting lead contacts the shell. The magnetic field is aligned along the wire axis. (b) Pair potential $\Delta$ and spectral gap $E_g$ as a function of the pair breaking parameter $\alpha$ in Maki's theory. As $\alpha$ goes to $\Delta_{00}/2$, superconductivity is ultimately destroyed (indicated by the vanishing pair potential), but prior to this the gap in the density of states vanishes. The symbols are estimated from tunneling conductance data reported in \cite{esteve2003} (triangles) and \cite{cm-hardgap} (squares). Below (b) is a plot of the same quantities for the purely Zeeman-driven transition, relevant to topological superconductivity.}
\label{fig:makisummary}
\end{center}
\end{figure}

The above two equations provide the content of the theory relevant to our discussion. In Fig.~\ref{fig:makisummary}(a) we plot the DOS calculated according Eq.~(\ref{eq:maki_eq2}) with the appropriate solutions $u$ from Eq.~(\ref{eq:maki_eq1}). We specifically choose values of $\alpha$ to correspond with the field values shown in Fig.~4a of \cite{cm-hardgap}, taking their reported value $B_c = 75$mT. We find, as shown in Fig.~\ref{fig:makisummary}(a), the same characteristic behavior (a broadening of the coherence peaks, introducing states below the bare pairing scale, but with a still well-defined gap feature) and even reasonable quantitative consistency with the experiment.

From the DOS we can read off another quantity, $E_g$, which is the measurable excitation gap. This is only equal to the pair potential in the BCS limit $\alpha = 0$. Otherwise the deviations are substantial, and indeed a finite region of ``gapless superconductivity" exists where $E_g = 0, \Delta \neq 0$. The expressions for $\Delta(\alpha)$ and $E_g (\alpha)$ at $T=0$ can be obtained analytically but are not very illuminating, so we do not reproduce them, however, they are plotted in Fig.~\ref{fig:makisummary}(b). In this figure, we also show (directly compared with the theoretical results) rough estimates of $E_g$ from the experiments reported in \cite{esteve2003} on a pure Al mesoscopic wire in a magnetic field, along with \cite{cm-hardgap}. Both experiments are done at very low but nonzero temperature, so the thermal ``tails" in the DOS make estimating $E_g$ at zero temperature a poorly defined task. Nonetheless, the general agreement of both experiments \cite{cm-hardgap, esteve2003} with the theory (and with one another) indicates that the magnetic field dependence of the conductance measurements of \cite{cm-hardgap} is dominated by the behavior of the Al shell.

To put this in the context of topological superconductivity, a well-known requirement for the topological quantum phase transition in the system is the closing (and then reopening) of the induced SC gap in the wire with increasing magnetic field purely through the Zeeman spin-splitting effect \cite{SauSM_PRB, StanescuSM_PRB}. The experimental data in Ref.~\cite{cm-hardgap}, in agreement with our theoretical results in Fig.~\ref{fig:makisummary} (and with the earlier experimental result for pure Al in Ref.~\cite{esteve2003}), show no spin splitting behavior in the magnetic field induced gap closing effect, which appears to be a purely pair-breaking orbital effect arising from the breaking of time reversal invariance in a small, disordered SC. In comparison to the Maki theory, below Fig.~\ref{fig:makisummary}(b) we show the same quantities for a clean SC in a Zeeman field without pair breaking, where the SC is eventually destroyed by Pauli paramagnetism \cite{clogston} \emph{well before} the excitation gap can close. For Al (with a $g$ factor of 2), and taking the value $\Delta_0 = 0.19$meV from \cite{cm-hardgap}, the Clogston critical field is $B_c = \frac{\Delta_0}{\sqrt{2} \mu_B} \simeq 2.32$T. The expected $g$ factor in InAs is substantially larger, $g \gtrsim 10$, so closing the \emph{induced} gap (equal here to $\Delta_0$) naively only requires a field $B \sim \frac{\Delta_0}{10 \mu_B} \simeq 0.33$T, safely below the Clogston limit of the Al shell, though still well above the experimental critical field (around $\sim 0.1$T in Ref.~\cite{cm-hardgap}). Explicitly, it is clear that the magnetic field dependence of the devices reported in \cite{cm-hardgap} comes purely from the orbital effect of the magnetic field, and surprisingly no spin physics is visible whatsoever at these fields. In the remaining sections we consider simplified models to understand and explain this result.


\section{Explicit modeling of the SM-SC interface}
\begin{figure}[tb]
\begin{center}
\includegraphics[width=0.48\textwidth]{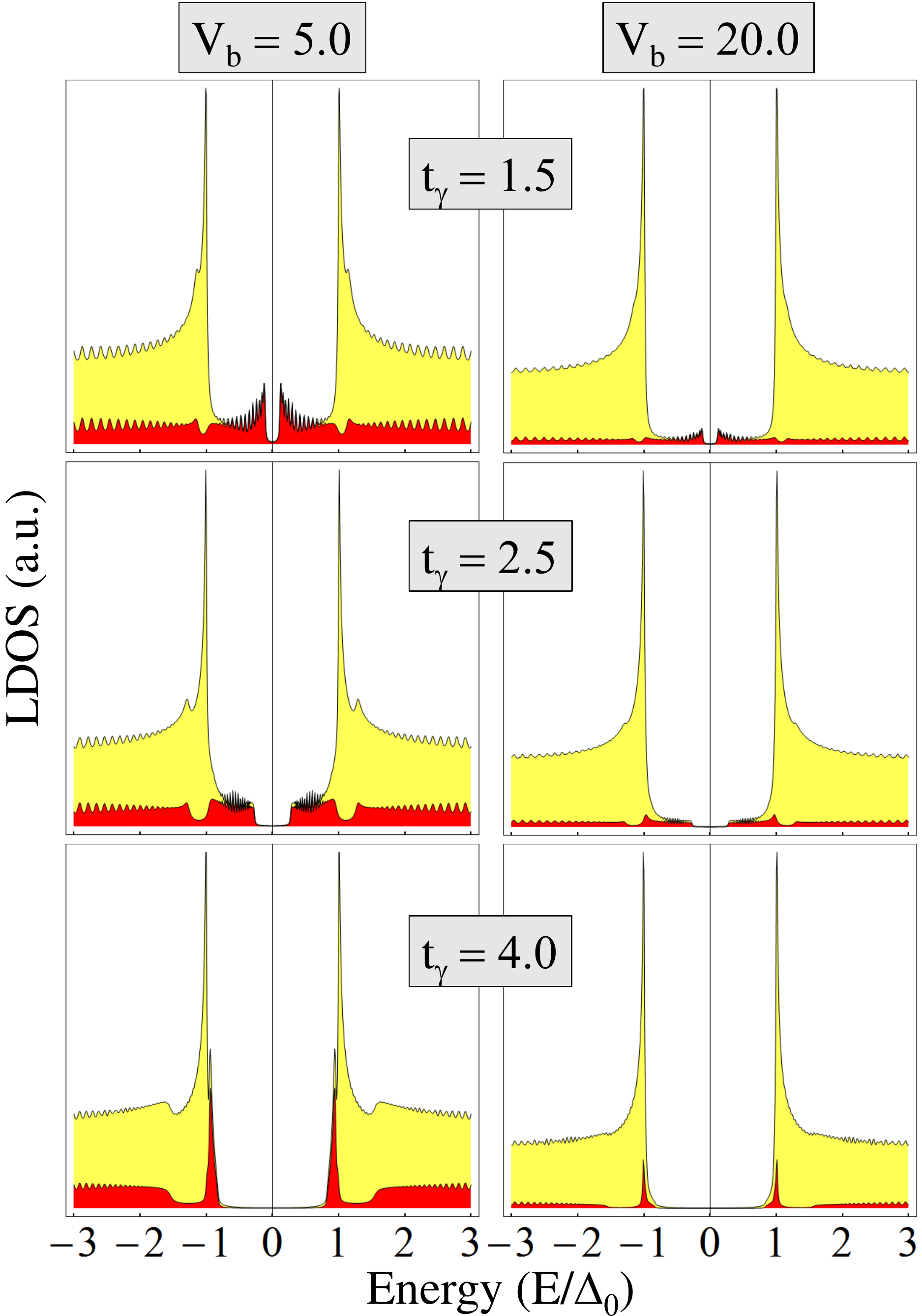}
\vspace{-4mm}
\end{center}
\caption{(Color online)  Local density of states integrated over a small region near the end of a SM wire - SC structure. The weight function decays exponentially on a scale $\delta_{SC}=8a$ and $\delta_{SM}=10a$, with $a$ the lattice spacing. The low-energy states are confined within the proximity-coupled structure by a Gaussian potential barrier of height $V_b$.  The SM-SC coupling is controlled by the hopping parameter $t_\gamma$. The values of the parameters are in units of $\Delta_0$. Red (dark gray) and yellow (light gray) correspond to contributions to the integrated LDOS from wire sites and SC sites, respectively. From top to bottom the panels correspond to weak, intermediate, and strong coupling regimes. Increasing the potential barrier height reduces the relative weight of the SM wire contributions, including the induced gap feature.}
\vspace{-3mm}
\label{Fig_LDOS}
\end{figure}

To clearly understand the implications of the non-local (short-range) tunneling scenario (i.e., to disentangle the SM and SC contributions) in the intermediate-to-strong coupling limit, we now calculate explicitly the LDOS for a proximity-coupled SM-SC structure using a simple multi-chain model. The semiconductor wire is modeled as a chain with nearest-neighbor hopping $t_{SM}=100\Delta_0$, where $\Delta_0$ is the bulk SC gap. The wire has a normal region, where a Gaussian barrier potential of height $V_b$ is applied. This potential confines the low-energy states inside a segment of length $N_{SM} = 600$ that is coupled to the bulk SC through the hopping matrix element $t_\gamma$, which controls the SM-SC coupling strength. The superconductor is modeled by six coupled chains of length $N_{SC} = 800$ with nearest-neighbor hopping $t_{SC}=10\Delta_0$. The chemical potentials are $\mu_{SC} = 36 \Delta_0$ and $\mu_{SM} = 0.5\Delta_0$ (as measured from the lowest energies of the uncoupled SC and SM spectra, respectively). These parameters ensure that the number of low-energy degrees of freedom of the SC is much larger than the number of SM degrees of freedom. The LDOS was integrated with a weight function that decays exponentially away from the potential barrier, so than only a few lattice sites from the SM wire and adjacent SC chain contribute significantly. The calculated LDOS results are shown in Fig.~\ref{Fig_LDOS}, illustrating how a typical tunneling conductance evolves from weak to strong coupling.

The red (dark gray) areas represent contributions due to the SM wire, while the yellow (light gray) areas are bulk SC contribution. We note two important features. First, the induced SC gap increases with the effective SM-SC coupling and becomes comparable with the bulk gap $\Delta_0$. As a result, the spectral features associated with the induced gap merge with the gap feature of the bulk SC.
Second, the relative weight of the SM wire contribution decreases with increasing barrier height. This behavior can be qualitatively understood in terms of the difference between the screening lengths in the SM wire and the bulk SC. Practically, increasing the barrier ``pushes" the SM states away from the end of the wire. On the other hand, strong screening makes this effect negligible in the bulk SC. 

Based on this numerical analysis, we arrive at the following qualitative physical picture concerning the information that can be extracted from a tunneling experiment on a hybrid SM-SC structure. i) In the weak tunneling (high barrier) regime -- the only regime that is consistent with the observation of a hard gap -- the conductance $dI/dV$ is dominated by features associated with the bulk SC (and not by the SM contributions, which are hardly visible in our numerical results). We note that the SM wire LDOS (red areas in Fig. \ref{Fig_LDOS}) is in fact strongly suppressed within a certain energy window right above the bulk gap $\Delta_0$. This is a generic feature of the SM wire LDOS; consequently, measured quantities (e.g., $dI/dV$) that lack this feature cannot be solely associated with the LDOS of the SM wire.  ii) To clearly distinguish the features associated with the wire (e.g., the induced gap) from bulk SC features, one must study their dependence on the potential barrier height. In particular for a half-shell InAs-Al device, which appears to correspond to intermediate coupling, the induced gap feature, clearly seen experimentally (in Fig. 5c of Ref.~\cite{cm-hardgap} as additional subgap structure similar to that in the middle-right panel of our Fig.~\ref{Fig_LDOS}), is expected to be reduced to arbitrarily low visibility relative to the bulk gap feature by increasing the height of the potential barrier. We note that in the case of a SM wire with multiband occupancy \cite{StanescuSM_PRB} the visibility of the induced features is strongly band-dependent. Specifically, increasing the barrier height will first suppress the features associated with the top occupied bands, particularly the Majorana band.   

One inevitable conclusion of our numerical results in Fig.~\ref{Fig_LDOS} is that the induced gap should be `large enough' for the intended application, while still maintaining its proximity-induced SM nature and not so large that the SC has overwhelmed all SM contributions. In addition to controlling the tunnel barrier, one must also have some way of reducing the strength of the SM-SC coupling (and consequently the size of the induced gap in the SM) which may happen naturally in the half-shell (or other fractional shell) hybrid structures since the contact area between the SM wire and the SC is automatically reduced in the fractional shell structures.


\section{Parameter renormalization}
To understand the SM states below the SC gap and how they approach the gap with increasing coupling, one can ``integrate out" the degrees of freedom of an assumed macroscopic superconducting reservoir \cite{jay-proxeffect,tudor-prox}. This leads to an expression for the Green's function describing the excitations of some generic system coupled to the reservoir
\be
G^{-1}(\omega) = \omega - H_0 - \Sigma(\omega)
\ee
where $H_0$ is the BdG hamiltonian of the system under consideration and
\be
\Sigma(\omega) = -\gamma \left( \frac{\omega + \Delta_0 \sigma_y \tau_y}{\sqrt{\Delta_0^2 - \omega^2}} \right)
\ee
is a self-energy resulting from uniform coupling of the system to the superconductor. To write this, we have adopted the 4-component Nambu spinor convention $(u_{\su}, u_{\sd}, v_{\su}, v_{\sd})$. We also note that the $g$ factor of the superconductor is presently taken to be zero for simplicity.

Although several authors have used this approach to construct effective low-energy models for topological superconductors with an assumed small $\gamma$, very few have emphasized the consequences of strong-proximity-coupling (where the assumption $\omega \ll \Delta_0$ must break down; see e.g., \cite{potter-eng, alicea-review, dassarma-njp, hoi-disorder-arxiv, peng-prl}). We demonstrate here that the renormalization of spin-splitting in the wire to its value in the superconductor can eliminate the topological phase transition completely by pushing the critical field in the wire above the Clogston limit of the parent superconductor. This implies that the $\gamma \gg \Delta_0$ limit must be avoided, and in fact allows one to estimate the maximum possible induced gap consistent with topological superconductivity in SM-SC systems. This is similar in spirit to the strong proximity coupling problem first pointed out in \cite{potter-eng} for 2D SC-SM interfaces: the Zeeman splitting in the SM is reduced with increasing induced superconducting gap, and therefore an optimal coupling must emerge to maximize the so-called ``topological gap" in the topological phase. However, there the Clogston limit of the parent superconductor was not taken into account.

\begin{figure}[t]
\begin{center}
\includegraphics[width=0.4\textwidth]{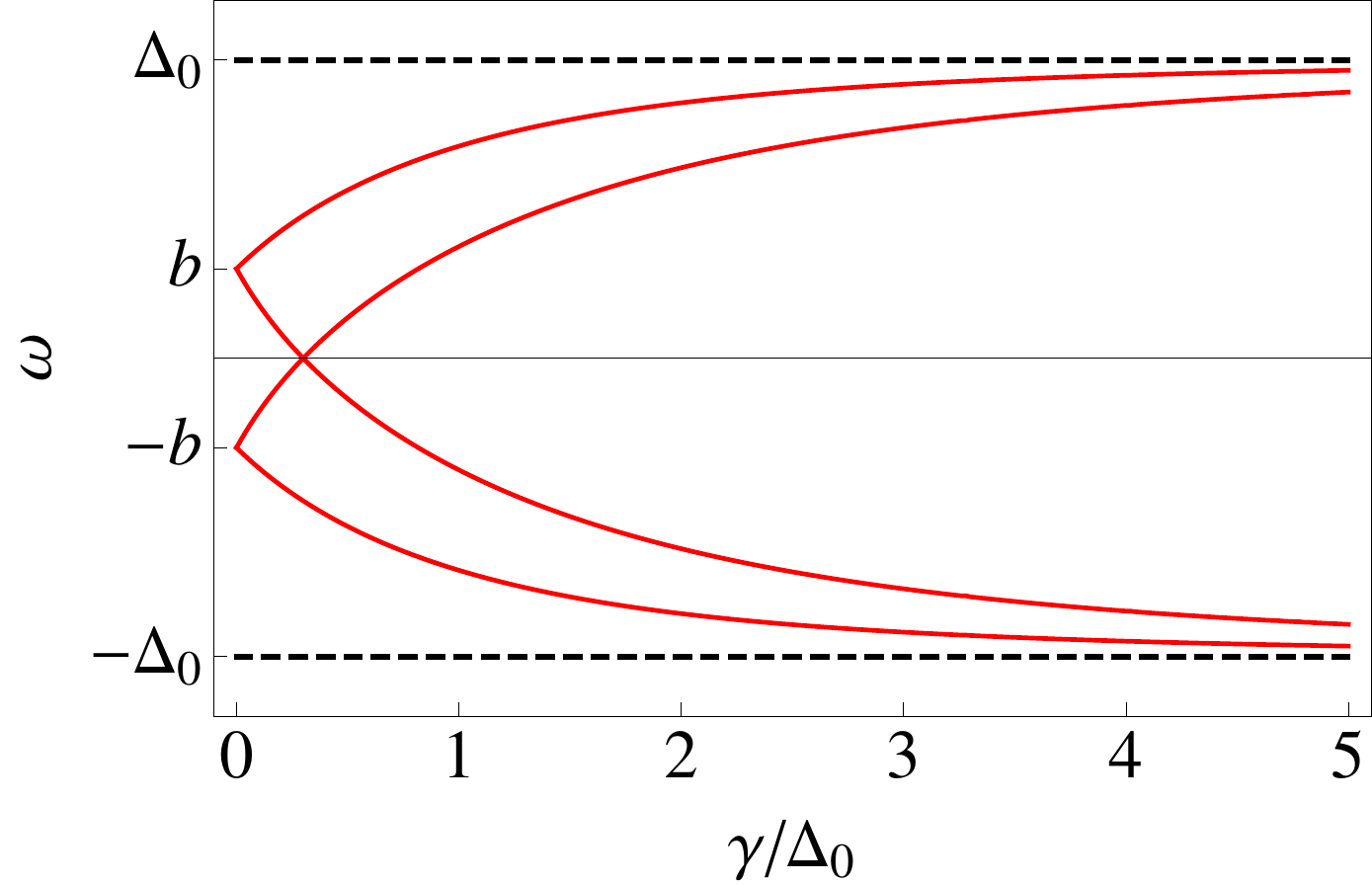}
\caption[Poles of the 0d single-particle Green's function]{(Color online)  Poles of the single-particle Green's function on the quantum dot, corresponding to spin-split Bogoliubov quasiparticle and quasihole states. For weak coupling to the superconductor, $\gamma/\Delta_0 \ll 1$, these appear near the energies associated with the regular spin-split states of $H_0$ at $\pm b$. As the coupling is made stronger, superconductivity dominates and the effect of the original spin splitting is weakened.}
\label{fig:0dbdgsolution}
\end{center}
\end{figure}

\begin{figure}[t]
\begin{center}
\includegraphics[width=0.50\textwidth]{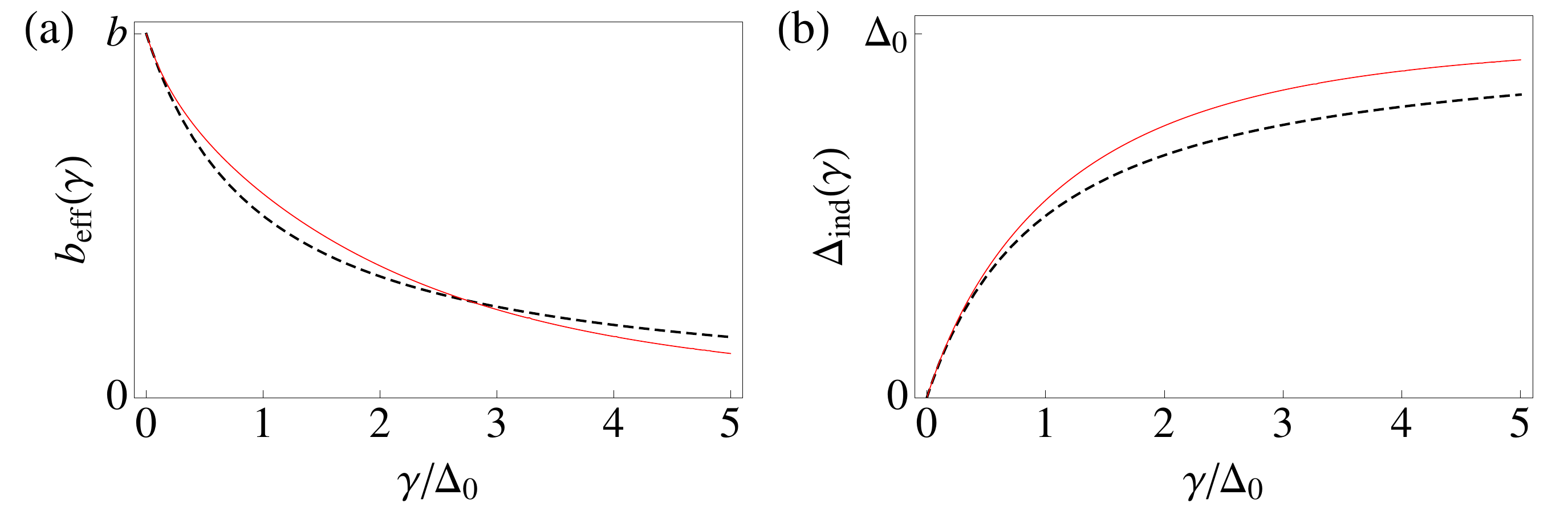}
\caption[Effective 0d hamiltonian parameters]{(Color online)  Effective hamiltonian parameters with increasing QD-SC coupling. (a) The effective Zeeman energy $b_{\rm eff}$ and (b) the induced pair potential $\Delta_{\rm ind}$. In both cases the solid red line is obtained from numerical solution of the BdG equation, while the dashed black line corresponds to the static approximation $\omega \rightarrow 0$.}
\label{fig:0drenormalization}
\end{center}
\end{figure}

We consider a minimal model of coupling to a single spin on a zero-dimensional quantum dot in a Zeeman field,
$H_0 = b \sigma_z \tau_z$,
where $b = \frac{g}{2} \mu_B B$ with $g$ the being the $g$-factor in the dot. This will be sufficient to demonstrate the renormalization of the spin splitting in different coupling regimes. We note as well that the model is very general, also describing, for example, a hamiltonian at fixed $k$ in an infinite wire or slab. We numerically obtain the solutions of $\Det{G^{-1}(\omega)} = 0$, yielding the spectrum shown in Fig.~\ref{fig:0dbdgsolution}. In the absence of coupling, the spectrum consists of spin up and down states separated by $2b$ (with the spectrum also doubled because of the particle-hole redundancy). As the coupling to the superconductor is increased, the redundant quasiparticle state moves from $-b$ to positive energy, so that at intermediate-to-strong coupling the spectrum simply looks like weakly spin-split superconducting quasiparticle states, corresponding to an effective BdG hamiltonian  $H(\gamma,\omega) = b_{\rm eff}(\gamma,\omega) \sigma_z \tau_z + \Delta_{\rm ind}(\gamma,\omega) \sigma_y \tau_y$, where $b_{\rm eff}(\gamma,\omega)$ and $\Delta_{\rm ind}(\gamma,\omega)$ have been substantially renormalized from their bare values. Following \cite{jay-proxeffect}, we can obtain the renormalization factors analytically,
\be
b_{\rm eff}(\gamma,\omega) = b \frac{\sqrt{\Delta_0^2-\omega^2}}{\sqrt{\Delta_0^2-\omega^2} + \gamma}
\ee
\be
\Delta_{\rm ind}(\gamma,\omega) = \Delta_0 \frac{\gamma}{\sqrt{\Delta_0^2-\omega^2} + \gamma}
\ee
and plot in Fig.~\ref{fig:0drenormalization} these quantities obtained from the numerical solution along with their corresponding approximations from taking the static limit $\omega \rightarrow 0$ (which clearly fails when $\gamma \gtrsim \Delta_0$).

On this basis, we can now make a semiquantitative argument for the implications for topological superconductivity to emerge in a more realistic nanowire system. In particular, the Zeeman energy in the wire has to be larger than the induced gap scale but below the Clogston limit in the parent SC. This requires satisfying the inequalities
\be
\frac{\Delta_{\rm ind}}{b_{\rm eff}} < 1 < \frac{\Delta_0}{\sqrt{2} b_{\rm sc}}
\label{eq:inequality}
\ee
with $b_{\rm sc} = b \cdot g_{\rm sc}/g_{\rm wire}$ the Zeeman energy in the bulk superconductor. Estimating from the above expressions in the static limit, we obtain
\be
\frac{\Delta_{\rm ind}}{b_{\rm eff}} = \left( \frac{\Delta_0}{\sqrt{2} b_{\rm sc}} \right) \frac{g_{\rm sc}}{g_{\rm wire}} \frac{\sqrt{2} \gamma}{\Delta_0}
\ee
Thus, satisfying the condition in Eq.~(\ref{eq:inequality}) becomes impossible (and therefore topological superconductivity and MZMs can never emerge in this system) unless $\gamma < \frac{g_{\rm wire}}{g_{\rm sc}} \frac{\Delta_0}{\sqrt{2}}$. This means that the SM-SC tunnel coupling (i.e. $\gamma$) must be in the weak or intermediate coupling regime -- that is, $\gamma$ of the same order as $\Delta_0$ (itself a typically small energy scale in absolute terms) -- which also sets a limit on the proximity-induced gap scale. For example, taking $g_{\rm wire}/g_{\rm sc} = 5$ leads to a \emph{maximum} proximity-induced gap in the nanowire which is only around 78\% of the parent SC gap. This is a direct conclusion of our theory with important implications for the on-going search for topological Majorana modes in hybrid SC structures.


\section{Experimental Implications}
Our theory provides specific suggestions for future designs using SM-SC hybrid nanowires to identify MZMs.

First, it is clear that having a hard induced gap in the nanowire is essential for the anyonic Majorana braiding (so as to minimize contamination from unwanted fermionic subgap states), and this breakthrough aspect of the epitaxial core-shell SM-SC materials \cite{krogstrup-natmat} must be maintained in all future Majorana systems.
Second, our work clearly establishes that the reported magnetic field dependent conductance data in Ref.~\cite{cm-hardgap} shows that a large induced gap in the semiconductor (comparable to that of the parent superconductor itself) may be highly detrimental to topological phenomena because the semiconductor nanowire may then manifest only the superconducting properties of the trivial parent material.
Third, this implies the conclusion that the hybrid material should \emph{not} be in the very strongly tunnel coupled SM-SC regime -- in fact, we explicitly derive the condition that the SM-SC tunnel coupling energy, while being uniform at the epitaxial interface, should be \emph{at most} on the order of the superconducting gap, which would then lead to a proximity-induced gap in the nanowire that is significantly less than the parent superconducting gap \cite{modelnote}.
 
An additional important and immediately experimentally verifiable prediction of our theory, which should be tested right away using the experimental set up and the existing full-shell and half-shell InAs/Al nanowire structures of Ref.~\cite{cm-hardgap}, is that the relative weight of the Al and InAs contributions to the experimentally measured differential conductance depends strongly on the tunnel barrier height  (i.e. on the pinch-off gate potential, not the back-gate potential in the set-up of Ref.~\cite{cm-hardgap}).  In particular, our theory makes a clear prediction (see Fig.~\ref{Fig_LDOS}, where the calculated LDOS is directly proportional to the experimental $dI/dV$ in a tunneling measurement) that decreasing the tunnel barrier height (i.e. $V_b$ in our Fig.~\ref{Fig_LDOS}) while keeping the SC-SM coupling fixed (i.e. keeping $t_\gamma$ fixed in Fig.~\ref{Fig_LDOS}) should lead to the subgap features contributed by InAs (below the parent Al superconducting gap) having a higher relative visibility.  This can be tested, for example, by measuring the $dI/dV$ for the half-shell nanowires (as in Fig. 5c of Ref.~\cite{cm-hardgap}) by varying the tunnel barrier height.  We predict a strong enhancement of the induced gap features as the tunnel barrier is decreased.  Such an experiment will directly confirm our theoretical finding that the strong coherence peaks in the experiment of Ref.~\cite{cm-hardgap} are coming from Al whereas the smaller subgap features originate in InAs.
 

\section{Conclusion}

We have addressed two related and important issues for strong-coupling hybrid SM-SC structures. The first question is what is actually being probed when measuring $dI/dV$ in epitaxially grown core-shell InAs/Al hybrid nanowires. We find that even qualitative consistency between theory and experiment demands that $dI/dV$ consists of contributions from both the SM wire and the bulk SC. Their relative strength depends on the strength of the potential barrier separating the system from the metallic lead. Establishing a \emph{hard} gap poses no challenge to producing MZMs, but requires a strong barrier which generally also strongly suppresses the relative contribution of the SM wire to the tunnel conductance. Future systematic experimental studies of $dI/dV$ as a function of the barrier height should help disentangle the SM wire and bulk SC contributions.
The second question is how optimal the large gap regime (obtained for full shell coating) is for observing MZM physics. Although strong coupling maximizes the induced gap (in fact, pushing it all the way up to the gap of the parent SC in Ref.~\cite{cm-hardgap}) it also renormalizes the key parameters of the wire. Above a certain critical value of the SM-SC coupling (and thus, a critical value of the induced gap in the nanowire) the topological SC phase becomes inaccessible. Future experiments might measure the critical field (as signaled by the zero-bias conductance peak) for different patterns and thicknesses of wire coverage, or introduce a tunnel barrier at the SM-SC interface (for example, by growing a thin insulating layer) to obtain a quantitative map of effective coupling strength.
Our work also indicates the need for additional control over the epitaxial InAs/Al interface so that the interface coupling, while being uniform to avoid effects of disorder, is not too large -- our theory predicts that this coupling should be on the order of the parent SC gap. This strict requirement on the coupling strength limits the maximum proximity induced gap (which is much more easily measured than the interface coupling strength itself) in the semiconductor to be significantly less than that in the parent superconductor for optimal Majorana hybrid SM-SC nanowires.

This work is supported by Microsoft and LPS-MPO-CMTC.


\input{proximity-revised-prb.bbl}
\bibliographystyle{apsrev4-1}

\end{document}

%% file: proximity-revised-prb.bbl
%

%% file: proximity-revised-prb.bbl
\begin{thebibliography}{50}
\bibitem [{mod()}]{modelnote}%
  \BibitemOpen
  \href@noop {} {}\bibinfo {note} {While these figures are based on a minimal
  model, and are not strictly universal, we expect them (as overall trends) to
  be fairly robust. The dependence on specific experimental details appears in
  the dependence of $b_{\rm eff}$ on $b$ which may generally include effects
  beyond our theory, such as magnetic flux focusing by the superconductor or
  modification of the wire parameters by interactions, etc.}\BibitemShut
  {Stop}%
\end{thebibliography}
